\documentclass[aps,prl,twocolumn,a4paper,10pt, longbibliography]{revtex4-1}

\usepackage[T1]{fontenc}		
\usepackage[english]{babel}		
\usepackage[latin1]{inputenc}	
\usepackage{times}

\usepackage{amsmath}			
\usepackage{amssymb}			
\usepackage{bbm}				
\usepackage{mathtools}
\usepackage{gensymb}
\usepackage{simpler-wick}
\usetikzlibrary{tikzmark}
\DeclarePairedDelimiterX\Dirbraket[3]{\langle}{\rangle}%
{#1\,\delimsize\vert\,\mathopen{}#2\,\delimsize\vert\,\mathopen{}#3}

\usepackage[colorlinks=true, a4paper=true, pdfstartview=FitV,linkcolor=blue, citecolor=blue, urlcolor=blue]{hyperref}

\usepackage{graphicx}			
\usepackage{graphics}			
\DeclareGraphicsExtensions{.pdf}
\usepackage{color}		

\newcommand{\bea}{\begin{eqnarray}}
\newcommand{\eea}{\end{eqnarray}}

\newcommand{\bk}{\mathbf{k}}

\begin{document}

\title{High-Temperature Phase Separation and Charge-Magnon Liquid in Kinetic Antiferromagnets}

\author{Johan~Carlstr\"om }
\affiliation{Department of Physics, Stockholm University, 106 91 Stockholm, Sweden}
\date{\today}

\begin{abstract}
Understanding mechanisms of quantum ordering in strongly correlated systems remains a central challenge in condensed matter physics, with implications for designing novel quantum materials. Here, we investigate kinetic antiferromagnetism on a triangular lattice under an applied magnetic field, where spin-polarons emerge as charge-magnon bound states with mutual attraction. Using large-scale diagrammatic Monte Carlo simulations, we show that this interaction drives high-temperature phase separation into charge- and magnon-rich regions, bordered by polarised Mott insulating voids. Spectral function analysis reveals a substantial energy correction from magnon interactions, indicating that these carrier-rich regions form a strongly bound charge-magnon liquid. These findings shed new light on recent experiments on MoTe$_2$/WSe$_2$ moiré bilayers, underscoring kinetic magnetism as a unique pathway for strong inter-carrier attraction and high-temperature quantum ordering, with potential applications in quantum materials.
\end{abstract}
\maketitle

The doped Mott insulator, with charge carriers propagating on a spin-background, is a paradigmatic model of strongly correlated electronic systems that is central for identifying unconventional pairing mechanisms \cite{RevModPhys.78.17}. In this context, two fundamental energy scales arise: The kinetic energy results from delocalisation of charge carriers, while the magnetic energy originates in super-exchange processes, which drive antiferromagnetism \cite{PhysRev.118.141,Polaron}. 

\begin{figure}[!ht]
\includegraphics[width=\linewidth]{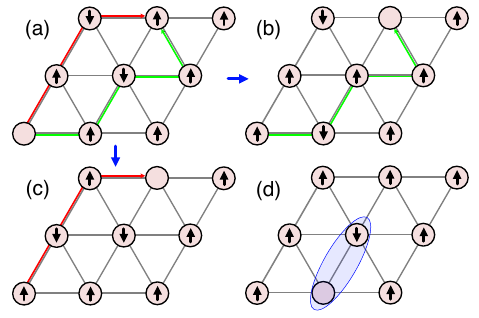}
\caption{
{\bf Kinetic antiferromagnetism} on the triangular lattice. In (a), the two paths lead to the same final position of the carrier. On a polarised background, these paths would interfere destructively, increasing the kinetic energy. However, in this initial configuration, the final states (b) and (c) are not identical, so no interference occurs. When destructive interference occurs, spin anti-correlation can reduce the kinetic energy, leading to antiferromagnetism. On a nearly fully polarised background, holes and magnons form strongly bound states, as shown in (d).
}
\label{KM}
\end{figure}

When the onsite repulsion greatly exceeds the bandwidth, super-exchange interactions are suppressed, making the kinetic energy scale dominant. In this scenario, magnetic order arises from delocalisation: on a polarised background, interference between carrier paths is much stronger than with anti-correlated spins. If this interference is constructive, delocalisation favours ferromagnetism; if it is destructive, it drives antiferromagnetism, as illustrated in Fig. \ref{KM}.

Kinetic magnetism has historically attracted attention in the context of He$_3$ crystals \cite{Andreev,Motambaux}. However, the advent of transition metal dichalcogenide (TMD) moiré heterostructures has dramatically expanded the possibilities for experimental realisation and control \cite{Zhang2020,PhysRevResearch.2.033087,PhysRevLett.121.026402,Xu2022}. While delocalisation generally promotes ferromagnetism on bipartite lattices \cite{PhysRev.147.392,PhysRevLett.108.126406}, TMD moiré materials enable low-energy degrees of freedom characteristic of a triangular lattice, supporting a wider range of behaviours, including antiferromagnetic interactions \cite{PhysRevLett.95.087202} and, in some cases, the ability to tune magnetic interactions in situ \cite{PhysRevResearch.4.043126}.

A central theoretical prediction of kinetic antiferromagnetism is that, under an external field, charges and magnons form bound states, including bipolarons that may condense to form a superconducting state. The binding energy for a single charge and a single magnon is estimated to be $\epsilon_{cm} \approx 0.4227t$ in the strong coupling limit, suggesting that these spin polarons are potentially very stable \cite{PhysRevB.97.140507}. At finite doping, the formation of a gas of such polarons is predicted to result in a magnetisation plateau at low temperatures \cite{PhysRevResearch.5.L022048}, accompanied by a pseudogap regime \cite{Zhang_2023}. In the partially polarised regime, magnons may mediate attractive interactions between charge carriers, leading to the formation of bound states \cite{PhysRevB.97.140507,PhysRevResearch.5.L022048,morera2021attractionfrustrationladdersystems}.

Experimental realisation of spin-polarons was recently reported in moiré bilayers \cite{Tao2024}. 
For moderate hole-doping, the magnetisation curve flattens out in the polaronic regime. This effect is only seen at low temperature ($T/t\sim 0.027$) where spin fluctuations remain small. 
When particle-doping the same material, the magnetic susceptibility is instead enhanced due to the formation of ferromagnetic polarons \cite{Ciorciaro2023}. 
With quantum gas microscopy, spin correlations in the proximity of carriers have been observed in both ferromagnetic and antiferromagnetic regimes \cite{Lebrat2024,Prichard2024,Xu2023}.

The recent advances in kinetic antiferromagnetism suggest a mechanism for generating attractive interactions between charge carriers and producing novel magnetic properties. However, the efficacy of this mechanism remains unclear, and its implications for macroscopic systems are still an open question, as prior studies have primarily focused on zero-temperature scenarios with limited numbers of charge carriers.

In this work, we exmine whether magnons can mediate significant attraction between charge carriers at finite temperatures, and, if so, explore the nature of the resulting phase and quantify the strength of these magnon-mediated interactions. By focusing on macroscopic systems with realistic thermal conditions, we aim to provide insights that bridge the gap between theoretical predictions and experimental realisations.

We tackle this problem using a minimalistic description of kinetic magnetism in the form of a triangular-lattice Hubbard model coupled to an external field
\bea
H\!=\!-\!\sum_{\sigma \langle ij\rangle} t c_{\sigma i}^\dagger c_{\sigma j} \!+\!\sum_i  \big[U\hat{n}_{\downarrow i} \hat{n}_{\uparrow i}+ B(\hat{n}_{\downarrow i}-\hat{n}_{\uparrow i})/2 \big]
\label{Hubbard}
\eea
in the strong-coupling limit $U/t\to\infty$. This scenario is relevant for existing moiré materials \cite{PhysRevResearch.2.033087,PhysRevLett.121.026402,Tang2020}, including the experiment  \cite{Tao2024} where it is estimated that the energy scales of hoping and super-exchange compare as $t/J\approx 20$.

\begin{figure}[!ht]
\includegraphics[width=\linewidth]{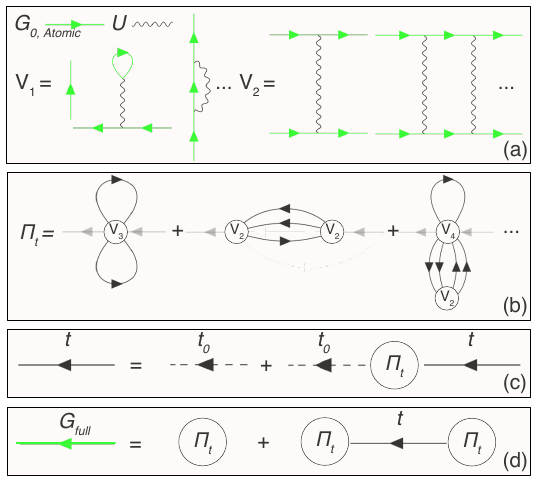}
\caption{
{\bf Strong coupling diagrammatic Monte Carlo} procedure. (a) The effective $n-$particle scattering vertices $V_n$ are computed from the bare local Greens function and the contact interaction $U$. This computation is conducted in the atomic limit, and can be done exactly and non-perturbatively, even for $U\to\infty$. (b) The polarisation operator of the hopping integral is obtained by from a diagrammatic expansion in $V_N$ and the dressed hopping integral $t$. This is done stochastically with some max order in $t$. (c) The dressed hopping integral is obtained from the polarisation operator via a Dyson/Bethe Salpiter type equation. The steps (b, c) thus form a closed set of equations which are iterated until convergence. (d) The full Greens function is obtained from the polarisation operator and the dressed hopping integral. 
}
\label{SCDMC}
\end{figure}

We address the model (\ref{Hubbard}) with diagrammatic Monte Carlo simulation, which is based on stochastic sampling of Feynman-type diagrams \cite{Van_Houcke_2010,PhysRevLett.119.045701,hou2024feynman,chen2024partial,PhysRevLett.121.130405}. To access the strongly correlated regime, we use a variation known as strong-coupling diagrammatic Monte Carlo (SCDMC), which relies on a self-consistent expansion in the dressed hopping integral, and was specifically developed for this scenario \cite{0953-8984-29-38-385602,PhysRevB.97.075119,PhysRevB.103.195147,carlstrom2021spectral,MAD}. The protocol, outlined in \cite{PhysRevB.103.195147}, is summarised in Fig. \ref{SCDMC}.

SCDMC is an asymptotically exact method in the sense that the only systematic source of error is truncation of the series. Furthermore, all results are obtained directly in the macroscopic limit, and for a convergent or resummable series, these can be obtained with known and typically small error bars. 

The principal observable produced in our simulations is the Gutzwiller-projected Greens function, defined on the permitted subspace \cite{PhysRevB.97.075119},
\bea
\tilde{G}_\sigma(x\!-\!x')\!=\!
\langle T [1\!-\!n_{-\sigma}(x')]  c_{\sigma}^\dagger(x')c_{\sigma}(x)[1\!-\!n_{-\sigma}(x)]  \rangle, \;\;\; \label{GGUTZ}
\eea
where $x,\;x'$ refer to points in space-time. From the Greens function, we derive two key observables, the equation of state and the quasiparticle spectrum. The former is obtained from noting that the carrier density is given by $-\tilde{G}(\mathbf{r}=0,\tau\to 0^+)$. The latter, denoted by $A(\bk,\epsilon)$, is related to the Greens function via
\bea
\tilde{G}(\bk ,\tau)=\int d\epsilon A(\bk,\epsilon) \frac{e^{-\epsilon \tau}}{1+e^{\beta \epsilon}}, \; \tau<0,\label{spectrum}
\eea
and can be obtained through simulated annealing 
\cite{PhysRevResearch.5.033160}. The total density of states is then obtained from
\bea
\text{d}(\epsilon)=\int \frac{d\bk}{(2\pi)^D}A(\bk,\epsilon). \label{dos}
\eea
The energy is discretised so that ($\epsilon\to \epsilon_i;\; 1\le i\le N_E$). We chose $N_E=121$ as a compromise between accuracy and numerical stability.

\begin{figure*}[!htb]
 \hbox to \linewidth{ \hss
\includegraphics[width=\linewidth]{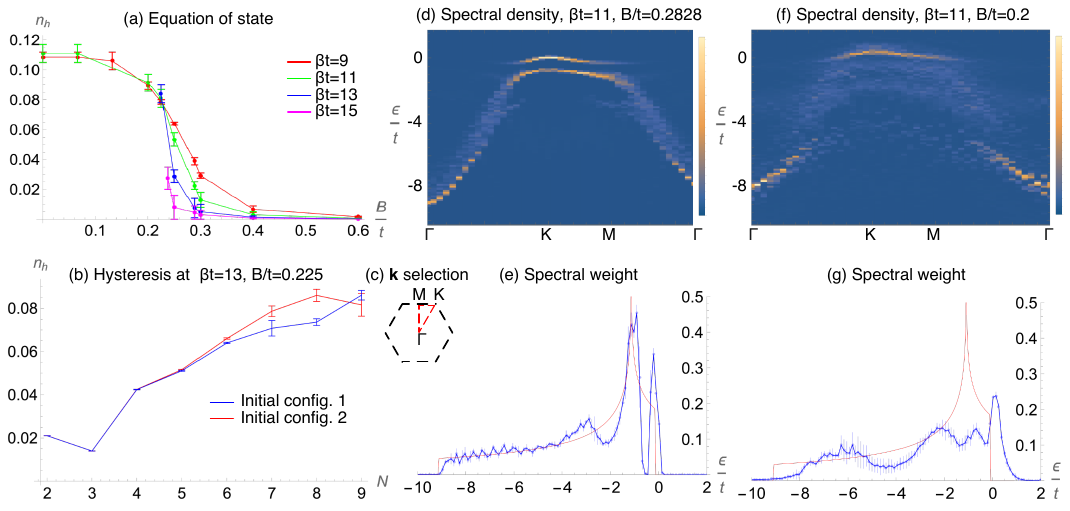}
 \hss}
\caption{
{\bf Results obtained} from strong coupling diagrammatic Monte Carlo. 
(a) Carrier density as a function of the external magnetic field $B$ with $\mu/t=3$ and inverse temperatures $\beta t=9,11,13,15$. Around $B/t\sim 0.25$ the carrier density declines rapidly with the magnetic field, revealing a large susceptibility which we interpret as resulting from phase separation. (b) Equation of state for $\beta t=13$ and $B=0.225t$ as a function of expansion order, obtained from two different initial configurations. At orders $N=7,8$ we observe that two solutions coexist, consistent with a phase separating system. 
(c) Illustrates the cut in momentum space used to plot spectral functions of the majority spin component. The spectra at $\beta t=11$ and $B/t=0.2818$ (d) reveals a "polaronic" sub-band of charges interacting with magnons. The corresponding spectral weight (e) indicates that this sub-band is gapped from the remaining spectral features. At $\beta t=11$ and $B/t=0.2$, the spectral function (f) becomes smeared, and the polaronic sub-band now overlaps with the rest of the spectrum (g). The spectral weight of fully polarised systems is shown in red (e, g). 
The width of the polaronic sub-band in (d) is estimated to $w_p\sim 0.5 t$, while it is notably broader in (f). 
}
\label{result}
\end{figure*}

The diagrammatic series must be truncated at some maximum order $N$, which is dictated by computational complexity. We are able to obtain solutions up to orders $N=7-9$, depending on model parameters. Our estimates for the equation of state are based on a progression of calculations where we monitor how the answer changes with the expansion order. A complete record of this process is provided in the Appendix.

In a self-consistent expansion, multiple solutions can theoretically be obtained if the free-energy landscape exhibits multiple minima. We primarily use solutions from lower expansion orders as initial guesses for the dressed hopping integral. At lower expansion orders, the solutions generally underestimate carrier density, so if multiple solutions exist, this approach tends to yield those with lower carrier concentration. 


Fig. \ref{result} summarises the results of the diagrammatic simulations. 
Panel (a) shows the equation of state as a function of the external field $B$. In the region $B/t\sim 0.25$, we observe a sharp drop in carrier density, which becomes more pronounced as the temperature decreases. 
At $\beta t=13$, most of the decline occurs within $B=0.225-0.25t$, a range of just $0.025t$.
 At $\beta t=15$, lower magnetic fields cannot be resolved due to convergence issues in the diagrammatic series, although the trend at stronger fields remains consistent. In the grand canonical ensemble, the presence of a large susceptibility, $\chi_{n,B}=\partial n/\partial B$, which increases at lower temperatures, is suggestive of phase separation.

To test the conjecture of phase separation, we use alternative initial configurations to identify signs of hysteresis. Fig. \ref{result} (b) provides two series computed at $\beta t=13, \; B/t=0.225$. The blue curve is obtained by using a solution from the previous expansion order, as explained above. For the red curve, a solution obtained at a weaker field, and thus greater carrier density, was used for initial condition. At orders $N=7,8$ we observe that this leads to an alternative solution. At order $N=9$, the first solution appears to becomes unstable. The difference in carrier density between these solutions is small, indicating that we are close to a bifurcation point where a second solution develops. At higher temperatures, we found that with alternative initial configurations, the systems converged to the same solution, but only very slowly, consistent with a flat free-energy landscape which would be expected close to a phase transition. 

The figures \ref{result} (d-g) show the spectral function and density of states at $\beta t=11$ with $B/t=0.2828$ (d-e) and $B/t=0.2$ (f-g) computed at an expansion order $N=7$, for the majority spin component. 
In the former case, the carrier density is small, at $n_h\sim 2\%$. The spectral density (d) reveals a polaronic sub-band that is separated from the remaining spectral features, consisting of carriers that reduce their kinetic energy by interacting with magnons in the environment. From the spectral weight (e) it is apparent that this sub-band is gapped from other degrees of freedom, consistent with the conjecture of a pseudo-gap phase in the polaronic metal \cite{Zhang_2023}. Is should be stressed that the gap is not situated at zero energy due to thermal spin and charge fluctuations. At zero temperature, we expect to see a gap at the fermi level.  
The width of the polaronic sub-band can be estimated to $w_p\approx 0.5 t$. 

At a weaker field of $B/t=0.2$, the carrier density increases to $n_h\sim 9\%$. The spectral function (f) now reveals a sub-band that is less well-defined, while remaining features are notably smeared. The density of states (g) shows that the polaronic sub-band is no longer gapped from the rest of the degrees of freedom. The width of the sub-band can for this reason not be completely made out, but a conservative estimate gives that $w_p>t$ in this case, and there is a notable tail at positive energy. 

The spectral function provides two essential implications for understanding polaronic metals. Firstly, the equation of state, Fig. \ref{result} (a), indicates that most of the carrier density vanishes within an interval of $\delta B=0.025t$. This corresponds to a change in polaron energy of $\delta \epsilon_p=3/2\; \delta B=0.0375t$ which is more than an order of magnitude smaller than the width of the sub-band. Thus, the observed susceptibility cannot be explained by flatness of the polaronic sub-band, providing further evidence for phase-separation. 

Secondly, at higher carrier densities, the sub-band width increases, implying that polarons hybridise. The highest-energy tail terminates at $\epsilon\approx t$, compared to the band top situated at $-0.1t$. This suggests a maximal carrier binding energy of $\epsilon_b\sim 1.1t$, notably larger than that of an isolated polaron. Since the corrections to single-particle energies from hybridisation is of a magnitude comparable to the polaron binding energy, these systems cannot be understood as a weakly interacting spin-polaron gas. We refer to this phase as a charge-magnon liquid (CML). 
The high onset temperature of phase separation ($T_c\sim t/13$), and the large shift in carrier energy sharply contrast the situation in doped Mott insulator on the square lattice, where inter-carrier attraction is barely discernible at comparable temperatures \cite{PhysRevResearch.3.013272}.

The emergence of a charge-magnon liquid should have major implications for the magnetic response, which are directly observable in experiments: In a gas of isolated polarons, spin-flips are associated with a single energy scale. This should result in a well-defined magnetisation plateau where the magnetic susceptibility vanishes \cite{PhysRevResearch.5.L022048}, provided that $ \epsilon_{cm}\gg T$. Also, the magnetic response should be almost temperature independent in this regime. 
By contrast, magnons in the CML phase should manifest across a continuum of energies. This scenario suggests that magnetic susceptibility will remain finite while displaying a distinct minimum, a behaviour that aligns with experimental observations. As doping levels increase, the observed susceptibility minimum rises, reflecting a trend that increasingly diverges from that expected in a polaron gas. The presence of a continuous range of magnon energies should also lead to thermal fluctuations that prevail at temperatures that are much smaller than the polaron binding energy, and this too is observed in the experiment \cite{Tao2024}. 

An additional observable relevant to the charge-magnon liquid is the integral over the second peak in the magnetic susceptibility, defined as
\bea
N_m=\int_{B_p}^\infty \chi_B dB, 
\eea
where $B_p$ represents the magnetic field at which the susceptibility $\chi_B$ reaches its minimum or plateau. This measure indicates the number of magnons remaining in the system upon reaching the plateau. Experimental data \cite{Tao2024} show that this quantity grows significantly with increasing doping, implying that carriers in the liquid state share excess magnons.
These observations clearly indicate that polaron hybridisation is relevant to real materials, including MoTe$_2$/WSe$_2$, despite the presence of long-range Coulomb repulsion.

In conclusion, we have investigated kinetic antiferromagnetism on a triangular lattice under an external magnetic field, focusing on magnon-mediated attraction between charge carriers. All the evidence point to the potency of this mechanism: the high-temperature onset of phase separation, the significant energy correction to carriers observed in the spectral function, and experimental indications of strong hybridisation all underscore the potential of kinetic magnetism as a pathway to high-temperature quantum-ordered states, including pairing. This behaviour contrasts sharply with Mott insulators on the square lattice and suggests that quantum gas spectroscopy \cite{BOHRDT2021168651, Gross995,Koepsell2019,koepsell2020microscopic} could reveal strong charge correlations, providing valuable insights into the nature of bound states. Additionally, our study highlights the need for ARPES experiments to estimate the energy scale of polaron hybridisation in real-world materials with Coulomb interactions, offering critical experimental benchmarks for our predictions.

Twisted transition metal dichalcogenides (TTMDs) present an especially promising platform for further exploration of kinetic magnetism, as they allow for continuous, in situ tuning of correlations between ferromagnetic and antiferromagnetic regimes \cite{PhysRevResearch.4.043126}, enabling precise control over magnon-mediated attraction. At low carrier density, with reduced attraction, hybridisation effects should be minimised, resulting in a gas of weakly interacting polarons and the emergence of a genuine magnetisation plateau. Ultimately, these insights could inform the design of quantum materials with tunable magnetic and electronic properties, advancing the broader understanding of unconventional magnetic states in strongly correlated systems.


This work was supported by the Swedish Research Council (VR) through grant 2018-03882. The computations were enabled by resources provided by the National Academic Infrastructure for Supercomputing in Sweden (NAISS), partially funded by the Swedish Research Council through grant agreement no. 2022-06725.

\bibliography{biblio.bib}

\section{Appendix}

\begin{figure*}[!htb]
 \hbox to \linewidth{ \hss
\includegraphics[width=\linewidth]{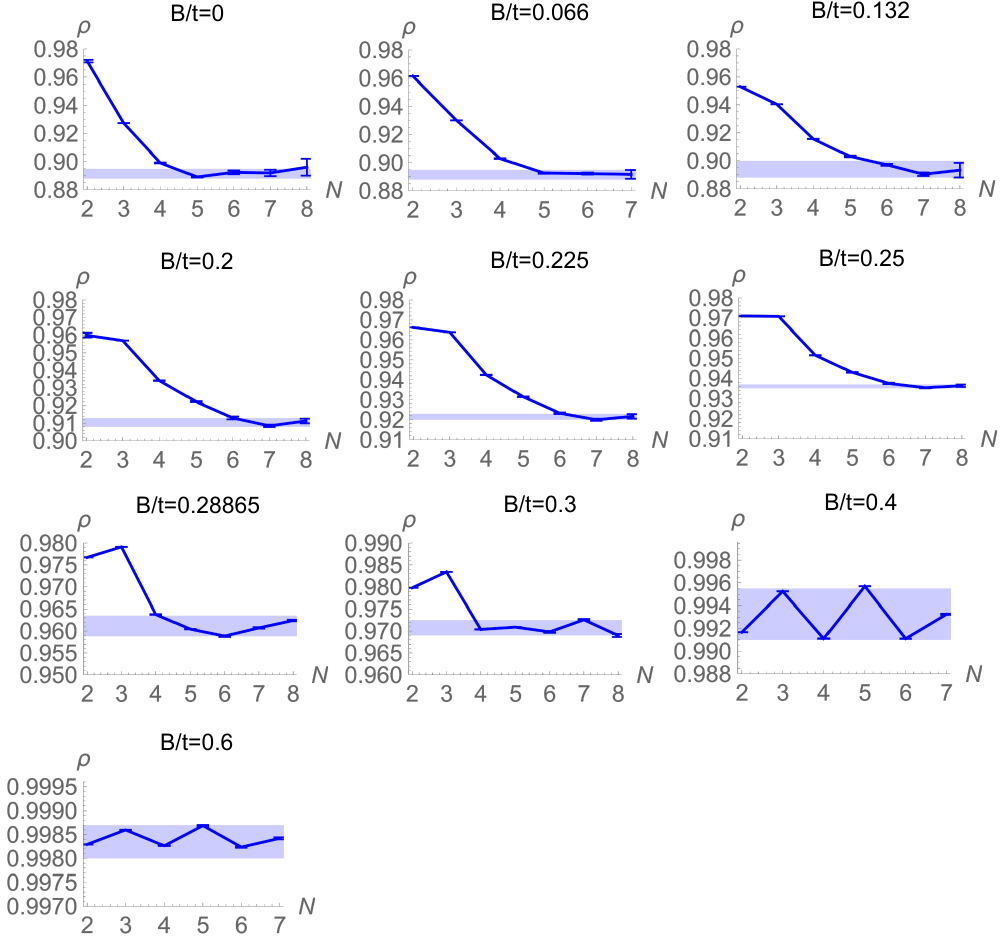}
 \hss}
\caption{
{\bf Equation of state} at $\beta t=9$ in an external magnetic field as a function of expansion order. 
}
\label{b9}
\end{figure*}

\begin{figure*}[!htb]
 \hbox to \linewidth{ \hss
\includegraphics[width=\linewidth]{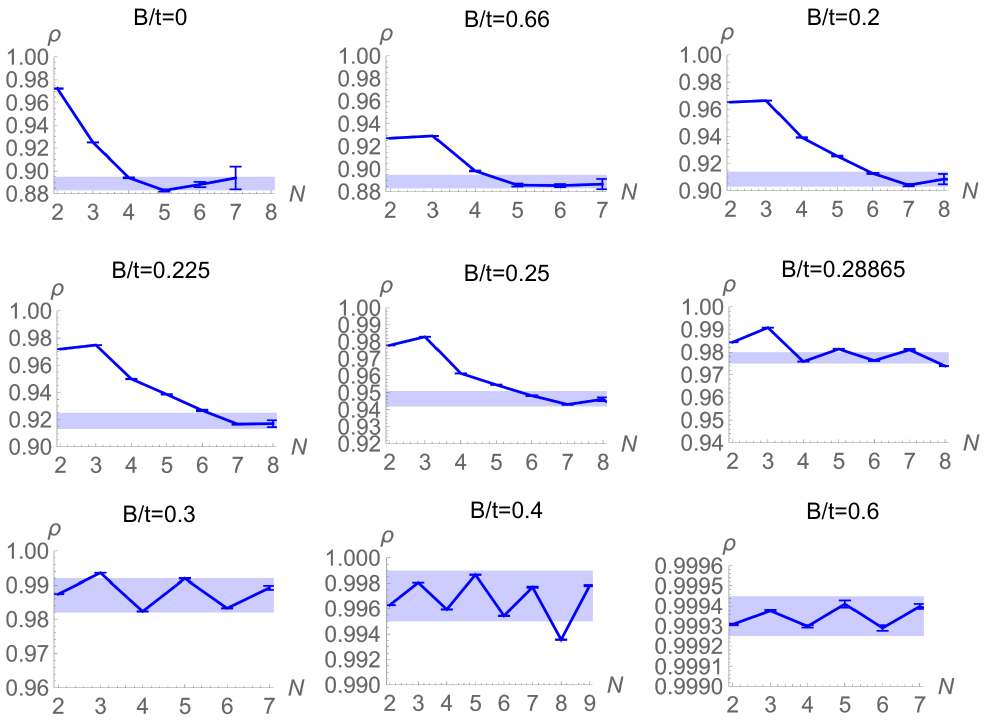}
 \hss}
\caption{
{\bf Equation of state} at $\beta t=11$ in an external magnetic field as a function of expansion order. }
\label{b11}
\end{figure*}

\begin{figure*}[!htb]
 \hbox to \linewidth{ \hss
\includegraphics[width=\linewidth]{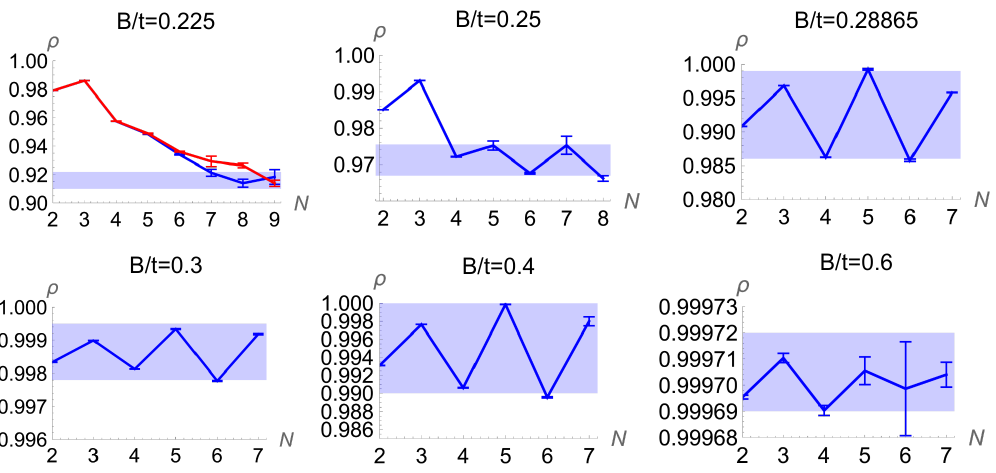}
 \hss}
\caption{
{\bf Equation of state} at $\beta t=13$ in an external magnetic field as a function of expansion order. At $B/t=2.25$, the system exhibits hysteresis at orders $7,8$, as revealed by altering the initial condition (ic) (Red: low carrier density ic. Blue: high carrier density ic). At order $N=9$ the solutions agree within error bars.  
}
\label{b13}
\end{figure*}

\begin{figure*}[!htb]
 \hbox to \linewidth{ \hss
\includegraphics[width=\linewidth]{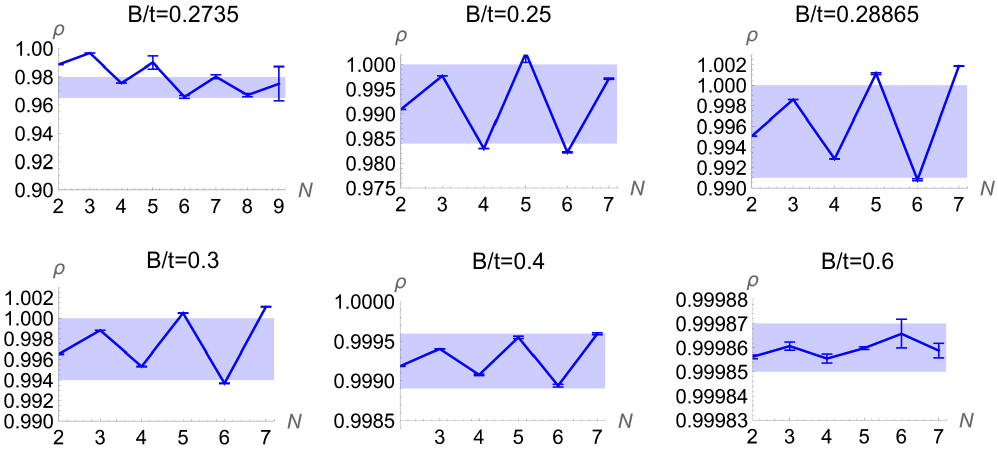}
 \hss}
\caption{
{\bf Equation of state} at $\beta t=15$ in an external magnetic field as a function of expansion order. At this temperature, we are limited to strong field and low carrier density. The series here appears to be asymptotic, consistent with a pole within the unit circle. 
}
\label{b15}
\end{figure*}

The figures \ref{b9}-\ref{b15} provide the equation of state obtained from the diagrammatic Monte Carlo simulations at temperatures $\beta t=9-15$, as a function of expansion order. The shaded area gives an estimate of the true particle density used in the main paper. At lower temperatures, we are often limited to lower carrier concentrations and thus a strong magnetic field. In these cases, the series typically oscillates, consistent with a pole situated inside the unit circle. 

Since we use a self-consistent method, it is principally possible to obtain multiple solutions. 
When conducting simulations, we generally use the solution obtained at a lower expansion order as a starting guess. At the lowest orders, the carrier density tends to be severely underestimated. Therefore, if multiple solutions exist, then this scheme tends to find the solution with the lower carrier concentration. To check if additional solutions exist, we also tried altering the initial configuration by using a solution obtained at a weaker field, which thus exhibits a higher concentration of charges. For $\beta t=13$ and $B/t=0.225$ we do indeed find alternative solutions at orders $N=7,8$ when feeding an initial configuration with a carrier density of $n_h\approx 10\%$. At order $9$, only the solution with a higher carrier concentration remains stable. 
At higher temperatures, we found that alternative initial configurations did not result in new solutions. However, we did find that convergence with respect to iterations was very slow. This is consistent with a flat free-energy landscape that would be expected before a second solution appears. 

\end{document}